\documentclass[preprint,showpacs,preprintnumbers,amsmath,amssymb]{revtex4}

\newcommand\rf[1]{(\ref{eq:#1})}
\newcommand\lab[1]{\label{eq:#1}}

\newcommand\br{\begin{eqnarray}}
\newcommand\er{\end{eqnarray}}
\newcommand\be{\begin{equation}}
\newcommand\ee{\end{equation}}

\newcommand\bc{\begin{center}}
\newcommand\ec{\end{center}}

\newcommand{\ct}[1]{\cite{#1}}
\newcommand{\bib}[1]{\bibitem{#1}}
\newcommand\PRL[3]{\textsl{Phys. Rev. Lett.} \textbf{#1}, #3 (#2)}

\newcommand\CMP[3]{\textsl{Commun. Math. Phys.} \textbf{#1}, #3 (#2)}
\newcommand\PRD[3]{\textsl{Phys. Rev.} \textbf{D#1}, #3 (#2)}

\newcommand\PTP[3]{\textsl{Prog. Theor. Phys.} \textbf{#1}, #3 (#2)}

\begin{document}

% \preprint{arxiv:[hep-th]}

\title{String Gas Shells, their Dual Radiation and Hedgehog Signature Control}

\author{E.I. Guendelman}%
\email{guendel@bgumail.bgu.ac.il}
% \author{E.I. Guendelman}%
% \email{guendel@bgumail.bgu.ac.il}
% \author{A. Kaganovich}%
% \email{alexk@bgumail.bgu.ac.il}
\affiliation{%
Department of Physics, Ben-Gurion University of the Negev \\
P.O.Box 653, IL-84105 ~Beer-Sheva, Israel}%

\begin{abstract}
We search for spherically symmetric, stationary solutions with a string gas shell as a source.
The requirement of a uniform newtonian potential, or constancy of the 00 component of the metric,
implies the existence of a  "dual" radiation, which we argue can be interpreted as representing the virtual quantum fluctuations
that stabilize the shell. A string hedgehog can be introduced also into the solution.
For zero or small hedgehog strength the string gas shell is of a regular nature, while the dual radiation
is of a spacelike nature. For higher hedgehog strengths however the radiation "materializes" and becomes timelike while the
string gas shell becomes space like. The significance of these solutions for the quantum theory is discussed.
\end{abstract}


\maketitle

%%%%%%%%%%%%%%%%%%%%%%%%%%%%%%%%%%%%%%%%%%%%%%%%%%%%%%%%%%%%%%%%%%%%%%%
\section{Introduction}

In recent investigations we have looked at the special role string gas shells play
in cosmology and in particular the possibility of constructing child universe solutions at 
an arbitrarily small energy cost \ct{child}. In this case the string gas shell is dynamical,
i.e., its radial coordinate is time dependent.

Here we want to see how solutions that involve a string gas shell can appear in a static case.
In this case, working with the assumption that
$ g_{00} $ is a constant, a very simple solution is obtained, where in addition to the string
gas shell a "dual" radiation must appear, this radiation appears in the simplest case as spacelike in nature.
This "dual radiation", we will argue, can be interpreted as representing the virtual quantum fluctuations
that stabilize the shell. 

The solutions can be generalized allowing for the introduction of a string-hedgehog \ct{string hedgehog} 
or a global monopole \ct{global monopole} on top of the string gas shell and its dual radiation. Then, for
big enough hedhehog strength, the solutions become cosmological, the radiation becomes timelike and the
string gas shell spacelike. In the situation of high hedgegog strength, the string gas shell could be 
interpreted as a quantum fluctuation of the string hedgehog configuration and that leads to the creation of 
the radiation in this case, as we will discuss.

%%%%%%%%%%%%%%%%%%%%%%%%%%%%%%%%%%%%%%%%%%%%%%%%%%%%%%%%%%%%%%%%%%%%%%%%%%%%
\section{Constant $ g_{00} $ solutions with string gas shell and its dual radiation}

Consider the following spherically symmetric distribution
\be
T^{\mu}_{\nu} = diag (-\rho, p_{r}, p_{\Omega}, p_{\Omega}) 
\lab{tensor energy momentum}
\ee

and the metric 

\be
ds^2 = - A dt^2 +  \frac{dr^2}{1-2m(r)/r} + r^2 d\Omega^2
\lab{metric}
\ee

then Einstein`s equations (we choose units where $G=c=1$) tell us then \ct{Grav. curv.},
\be
\frac{dm(r)}{dr}= 4 \pi r^2 \rho
\lab{mass function}
\ee

\be
 \frac{1}{2A} \frac{dA}{dr}= \frac{m(r) + 4 \pi p_{r} r^3}{r(r-2m)}
\lab{A gradient}
\ee

\be
\frac{d p_{r}}{dr} = - \frac{1}{2A} \frac{dA}{dr} (\rho + p_{r})+  \frac{2}{r} (p_{\Omega}-p_{r})
\lab{pr gradient}
\ee

We now require to have a shell of matter, i.e. 

\be
\rho = \sigma \delta(r-a)
\lab{shell}
\ee

here $\sigma$ is a constant representing the surface energy density in the shell. In addition to this condition
we require the additional requirement of a constant $ g_{00} < 0 $ , which by a rescaling of the time
coordinate can be set to $ g_{00} = -1 $. Since $A = constant $, we obtain from \rf{A gradient}

\be
p_{r}= - \frac {m(r)}{4 \pi  r^3} 
\lab{pr equation}
\ee

The condition $ g_{00} = constant $ was postulated also to specify solutions where a smooth gaussian type
distribution for the energy density (motivated from non commutative geometry) was used instead of \rf{shell} in \ct{Spallucci}.

Combining \rf{pr equation} with \rf{pr gradient} and \rf{mass function}, we obtain,

\be
p_{\Omega}= - \frac {1}{2} \rho + \frac {m(r)}{8 \pi  r^3} 
\lab{pomega equation}
\ee

We now consider the distribution given by \rf{shell}, which gives from \rf{mass function} that $m(r) = 4 \pi  \sigma a^2 \theta(r-a)$
if we take as a boundary condition that $m(0) = 0$ in order not to have singularities in the center of the geometry.
Therefore from \rf{pr equation} and \rf{pomega equation} we obtain,

\be
p_{\Omega}= - \frac {1}{2} \sigma \delta(r-a) + \frac {\sigma a^2 \theta(r-a)}{2 r^3} 
\lab{strig plus rad 1}
\ee

\be
p_{r}= - \frac { \sigma a^2 \theta(r-a)}{r^3} 
\lab{string plus rad 2}
\ee

The relation between the energy density, which is a delta contribution and the delta type contribution to the pressure $p_{singular}$ , or the shell contribution, is $p_{singular} =  - \frac {1}{2} \rho $, exactly like the
equation of state of a string gas in two spatial dimensions, which is in fact of the form $p =- \frac {1}{2} \rho$.

The theta type contribution is totally spatial,
contributing to pressure and not to the energy density. Furthermore, the sum of the pressures 
 of the theta type contributions ( that is their contribution to $p_{r} + 2p_{\Omega}$) is
zero, so the theta type contribution gives us a traceless piece to the energy momentum tensor and in this sense we
call it the "dual " radiation to the string gas shell. 

It is crucial to notice that the radial pressure outside the shell  \rf{string plus rad 2} is negative, which is the reason why the outside stress
provides a stabilizing effect: without this the string gas shell naturally tends to collapse to the center of the geometry, but the situation where the outside radial
pressure being negative, while the inside radial pressure is zero causes a pressure difference that pushes the string gas in the outside direction,
which we see, exactly equilibrates the effect of the tension of the strings that pushes the strings towards collapse.

We can  interpret the dressing of the string shell gas with these
theta type contributions, that are required to get the static self consistent shell solution, as representing the virtual quantum fluctuations
that stabilize the shell. Without this additonal stress, the string gas shell will simply collapse, but we can assume that there will be quantum effects that stabilize the string gas shell. What we have found phenomenologically is an object that does the same stabilizing
job as the quantum fluctuations are supposed to do. It is reasonable then to take an additional step and think of the theta type of contributions
as indeed being a classical representation of the virtual quantum fluctuations that stabilize the string gas shell. The detailed analysis of the
quantum theory will probably lead to some additional information, like for example a relation between $a$ and $\sigma$, or  may be some other condition (or restriction).

As we will see in the next section, by the introduction of a hedgehog of strings of big enough strength, the "dual" radiation can become time like, while the string gas shell becomes spacelike, that is there is the possibility that the virtual fluctuations that the theta terms represent "materialize". This lends further support to this interpretation.

%%%%%%%%%%%%%%%%%%%%%%%%%%%%%%%%%%%%%%%%%%%%%%%%%%%%%%%%%%%%%%%%%%%%%%%%%%%%
\section{Introduction of the Hedgehog}
A hedgehog of strings \ct{string hedgehog} is a spherically symmetric ensemble of strings emanating from the origin and 
pointing radially to infinity. It produces an energy momentum of the form

\be
T^{\mu}_{\nu} = diag (-\rho, p_{r}=-\rho, p_{\Omega}=0, p_{\Omega}=0) 
\lab{hedgehog}
\ee
This form for the energy momentum tensor is also satisfied for a global monopole \ct{global monopole}.

The formalism considered here, with constant $A$, which leads to the eqs. \rf{pr equation} and \rf{pomega equation}
allow us to generate also a string hedgehog provided we modify our assumption for the energy density and take now instead of
 \rf{shell}

\be
\rho = \sigma \delta(r-a) + \frac{h}{r^2}
\lab{shell plus hedgehog}
\ee

then the application of eqs. \rf{mass function}, \rf{pr equation} and \rf{pomega equation}
lead now to
\be
m(r) = 4 \pi  \sigma a^2 \theta(r-a) +  4 \pi h r
\lab{with hedgehog 1}
\ee

once again, as in the previous section, assuming the boundary condition  $m(0) = 0$ 

\be
p_{r} = - \frac{h}{r^2}- \frac { \sigma a^2 \theta(r-a)}{r^3} 
\lab{with hedgehog 2}
\ee

\be
p_{\Omega} = - \frac{1}{2}\sigma \delta(r-a) + \frac { \sigma a^2 \theta(r-a)}{2r^3} 
\lab{with hedgehog 3}
\ee

Notice that the calculation does not give  h-contributions to $p_{\Omega}$ and the h-contribution
to $p_{r}$ is exactly minus the h-contribution to the energy density, therefore confirming that the
h-contribution is of the form \rf{hedgehog}, that is consistent with that of a string hedgehog 
(or alternatively a global monopole).

We see that if $h > \frac{1}{8\pi}$ the r-r component of the metric becomes negative everywhere,
so that r becomes a time like coordinate. In order to have a Minkowskii signature space, the constant $A$
must be taken to be negative, so that now t is a space like coordinate.

As a consequence, the dual radiation becomes now time like evolving in time r and being
diluted as $\frac{1}{r^3}$ consistent with dilution of radiation in the cosmological expansion of the $\Omega$ sphere
which expands like $r^2$. The string gas shell exists now only for a moment in time $r=a$ and from that time on
it is substituted by the dual radiation, now of time like nature.

%%%%%%%%%%%%%%%%%%%%%%%%%%%%%%%%%%%%%%%%%%%%%%%%%%%%%%%%%%%%%%%%%%%%%%%%%%%%
\section{Discussion and Conclusions}
In this paper we have studied the gravitational field that is produced by a string gas shell plus an
associated dual radiation determined by the condition that $ g_{00} $ is a constant.

We can  interpret the dressing of the string shell gas with these
dual radiation contributions, that are required to get the static self consistent shell solution, as representing the virtual quantum fluctuations
that stabilize the shell. Without these additional stress, the classical string gas shell will simply collapse, but we can assume that there will be quantum effects that stabilize the string gas shell. What we have found phenomenologically is an object that does the same stabilizing
job as the quantum fluctuations are supposed to do, it is reasonable then to take the additional step and think of the dual radiation type of contributions as indeed being a classical representation of the virtual quantum fluctuations that stabilize the string gas shell.

Also the effects of a string hedgehog can be incorporated and the hedgehog has the potential of "materializing" these virtual contributions.

For low or zero values of the string hedgehog strength $h$, the string gas shell is of a regular nature, 
while the dual radiation is of a spacelike nature, however when the hedgehog strength is high enough, r becomes 
a time like coordinate everywhere and $-p_{r}$ has now the interpretation of an energy density which is non zero for the
dual radiation. The string gas shell by contrast has $p_{r}=0$, i.e. no temporal component for its energy 
momentum tensor in this regime. The string gas shell exists now only for a moment of time $r=a$, it may be interpreted 
as a quantum fluctuation (which exists for a very short time) of the strings producing the string hedgehog but that leads
to the generation of the radiation that exists for $r>a$. 
 
The strength of the hedgehog appears to control then which component of the energy momentum tensor is "virtual" or space like 
and which has "materialized" as real particles.

This conversion of "virtual excitations" into real ones, appears similar to the role external fields can have in converting
virtual particles into real ones as it is known to take place in the Schwinger mechanism \ct{Schwinger} in the presence of an external electric field
or the Hawking mechanism \ct{Hawking} in the vecinity of a black hole. 

%%%%%%%%%%%%%%%%%%%%%%%%%%%%%%%%%%%%%%%%%%%%%%%%%%%%%%%%%%%%%%%%%%%%%%%%%%%%

%\vspace{-0.5cm}
\section*{Acknowledgments}

% \vspace{-0.5cm}
I would like to thank Euro Spallucci for very usefull conversations,
G.Cantatore for hospitality at the University of Trieste and INF Trieste
for support.

%%%%%%%%%%%%%%%%%%%%%%%%%%%%%%%%%%%%%%%%%%%%%%%%%%%%%%%%%%%%%
% Doing references:                                         %
%%%%%%%%%%%%%%%%%%%%%%%%%%%%%%%%%%%%%%%%%%%%%%%%%%%%%%%%%%%%%

\end{document}